\documentclass[prl,twocolumn,superscriptaddress,showpacs]{revtex4}

\usepackage{graphicx}
\usepackage{dcolumn}
\usepackage{bm}

\begin{document}

\preprint{ATB-1}

\title{Ferromagnetic in-plane spin fluctuations in Na$_x$CoO$_2$ observed by neutron inelastic scattering}

\author{A.T. Boothroyd}
\email{a.boothroyd1@physics.ox.ac.uk}
\homepage{http://xray.physics.ox.ac.uk/Boothroyd}\affiliation{
Department of Physics, Oxford University, Oxford, OX1 3PU, United
Kingdom }
\author{R. Coldea}\affiliation{
Department of Physics, Oxford University, Oxford, OX1 3PU, United
Kingdom }
\author{D.A. Tennant}\affiliation{
School of Physics and Astronomy, University of St Andrews, St
Andrews, Fife, KY16 9SS, United Kingdom }
\author{D. Prabhakaran}\affiliation{
Department of Physics, Oxford University, Oxford, OX1 3PU, United
Kingdom }
\author{L.M. Helme}\affiliation{
Department of Physics, Oxford University, Oxford, OX1 3PU, United
Kingdom }
\author{C.D. Frost}\affiliation{
ISIS Facility, Rutherford Appleton Laboratory, Chilton, Didcot,
OX11 0QX, United Kingdom }
\date{\today}

\begin{abstract}
We present neutron scattering spectra taken from a single crystal
of Na$_{0.75}$CoO$_2$, the precursor to a novel cobalt-oxide
superconductor. The data contain a prominent inelastic signal at
low energies ($\sim 10$\,meV), which is localized in wavevector
about the origin of two-dimensional reciprocal space. The signal
is highly dispersive, and decreases in intensity with increasing
temperature. We interpret these observations as direct evidence
for the existence of ferromagnetic spin fluctuations within the
cobalt-oxygen layers.
\end{abstract}

\pacs{75.40.Gb, 74.25.Ha, 74.20.Mn, 78.70.Nx}
\maketitle

Sodium cobalt oxide (Na$_x$CoO$_2$) has become of considerable
interest in the past few years owing firstly to the report of a
large thermoelectric power coupled with low resistivity in single
crystals of Na$_{0.5}$CoO$_2$ \cite{Terasaki-PRB-1997}, and
second, to the serendipitous discovery of superconductivity at
temperatures below $T_{\rm c} \simeq  5$\,K in Na$_x$CoO$_2 \cdot
y$H$_2$O ($x \simeq 0.3$, $y \simeq 1.3$) formed by hydration of
precursor Na$_{x}$CoO$_2$ \cite{Takada-Nature-2003}. The enhanced
thermoelectric properties of Na$_x$CoO$_2$ make it potentially
important for thermoelectric applications, and the existence of
superconductivity in the hydrated compound raises very interesting
questions about the pairing mechanism in relation to the cuprate
and ruthenate superconductors.

Like the superconducting cuprates and ruthenates, the structure of
Na$_x$CoO$_2$ is layered, consisting in this case of triangular
CoO$_2$ sheets separated by layers of Na$^+$ ions. Incorporation
of water increases the sheet separation and decreases the Na
content. Metallic behaviour is achieved by doping the
Mott-insulating CoO$_2$ sheets with electrons donated by Na, which
create Co$^{3+}$ ions with zero spin in a background of Co$^{4+}$
carrying spin $S=\frac{1}{2}$. The enhanced thermopower has been
ascribed to a large spin entropy arising from strong
electron--electron interactions in the CoO$_2$ sheets
\cite{Wang-Nature-2003}.

Superconductivity in the hydrated compound is observed in a narrow
range of composition centred around $x = 0.3$
\cite{Schaak-Nature-2003}. Although the pairing mechanism has not
yet been established, a number of experimental studies have
uncovered evidence suggesting that the superconductivity in this
material is unconventional \cite{Unconventional-SC}. Several
attempts
\cite{Baskaran-PRL-2003,Kumar-Shastry-PRB-2003,Ogata-JPSJ-2003,Honekamp-PRB-2003,Wang-cm0304377}
to understand the superconductivity have been based on Anderson's
resonating valence bond (RVB) idea \cite{Anderson-Science-1987},
which describes a singlet quantum spin liquid ground state
stabilized by frustration. The $S=\frac{1}{2}$ triangular lattice
could in principle support an RVB state providing the coupling
between nearest-neighbour spins were antiferromagnetic, as
suggested by the high temperature susceptibility of both
superconducting Na$_{0.35}$CoO$_2 \cdot 1.3$H$_2$O
\cite{Sakurai-cm0304503-2003} and precursor Na$_x$CoO$_2$
\cite{Ray-PRB-1997}.
Among alternative scenarios, the possibility of spin-triplet
superconductivity has been analyzed by several authors
\cite{Baskaran-PRL-2003,Tanaka-Hu-JPSJ-2003,Ikeda-JPSJ-2003,Tanaka-JPSJ-2003},
and has received support from resonance experiments
\cite{Unconventional-SC}. The existence of ferromagnetic
correlations, which would open the door to $p$-wave spin-triplet
pairing, was predicted from density-functional calculations for
Na$_{x}$CoO$_2$ with $x=0.3$--0.7
\cite{Singh-PRB-2000,Singh-PRB-2003}. The predicted ferromagnetic
instability was found to be robust to the structural changes
associated with hydration \cite{Singh-PRB-2003}. Proximity to a
ferromagnetic instability is suggested by an irreversible magnetic
transition at $T_{\rm m} = 22$\,K in Na$_x$CoO$_2$
\cite{Motohashi-PRB-2003}, an upturn in the low temperature
susceptibility of Na$_x$CoO$_2$
\cite{Motohashi-PRB-2003,Prabhakaran-2003} and Na$_{0.35}$CoO$_2
\cdot 1.3$H$_2$O \cite{Sakurai-cm0304503-2003}, and has also been
inferred from some nuclear relaxation measurements
\cite{Unconventional-SC}.

The importance of spin degrees of freedom, both to explain the
thermopower and in most of the theories of superconductivity
proposed so far, makes it is essential to have experimental
information on magnetic correlations, especially to establish
whether the dominant in-plane coupling is ferromagnetic or
antiferromagnetic. This information can most directly be obtained
from neutron scattering measurements on single crystal samples.
Crystals of superconducting compositions large enough for neutron
scattering are not yet available. Here we present the first
neutron inelastic scattering spectra of the precursor compound
Na$_{x}$CoO$_2$. The data provide strong evidence for the
existence of ferromagnetic spin correlations within the CoO$_2$
layers.

Single crystals of Na$_{0.75}$CoO$_2$ were grown by the
floating-zone method in an image furnace \cite{Prabhakaran-2003}.
A crystal of size $\sim 10\times 8 \times 3$\,mm$^3$ was cleaved
from the zone-melted rod. Magnetization and powder X-ray
diffraction measurements made on samples taken from adjacent parts
of the same crystal rod showed no impurity phases within the
detectable limit of $\sim 2$\,\%.

Neutron inelastic scattering measurements were made on the MAPS
spectrometer at the ISIS Facility. MAPS is a time-of-flight
instrument equipped with a large pixellated detector surrounding
the incident beam direction. Neutrons of well-defined incident
energy are delivered to the sample in short pulses. Spectra in
each pixel are recorded as a function of neutron time-of-flight,
and subsequently transformed into an intensity array in $({\bf
Q},E)$ space, where ${\bf Q} = {\bf k}_{\rm i} - {\bf k}_{\rm f}$
is the scattering vector (the difference between the incident and
final neutron wavevectors) and $E$ is the energy transferred to
the sample. The intensity was converted into an absolute cross
section by comparison with the scattering from vanadium. The
presented spectra are the partial differential cross section ${\rm
d}^2 \sigma/{\rm d}\Omega{\rm d}E_{\rm f}$ per formula unit (f.u.)
multiplied by the factor $k_{\rm i}/k_{\rm f}$ \cite{Squires},
where $E_{\rm f}$ is the final energy.

The crystal was attached to a copper rod and mounted on a
closed-cycle cooler. Measurements were made at several
temperatures between 6\,K and room temperature. Typical counting
times were 36\, hours at an average proton current of 170\,$\mu$A.
An initial examination by neutron Laue diffraction showed that the
crystal contained several grains with an overall mosaic spread of
$\sim 10$ degrees. To maintain good $\bf Q$ resolution we employed
relatively low incident energies ($E_{\rm i}=40$\,meV and
60\,meV), and restricted our measurements to small scattering
angles.

\begin{figure}
\begin{center}
\includegraphics
[width=8cm,bbllx=105,bblly=314,bburx=462,
bbury=757,angle=0,clip=]{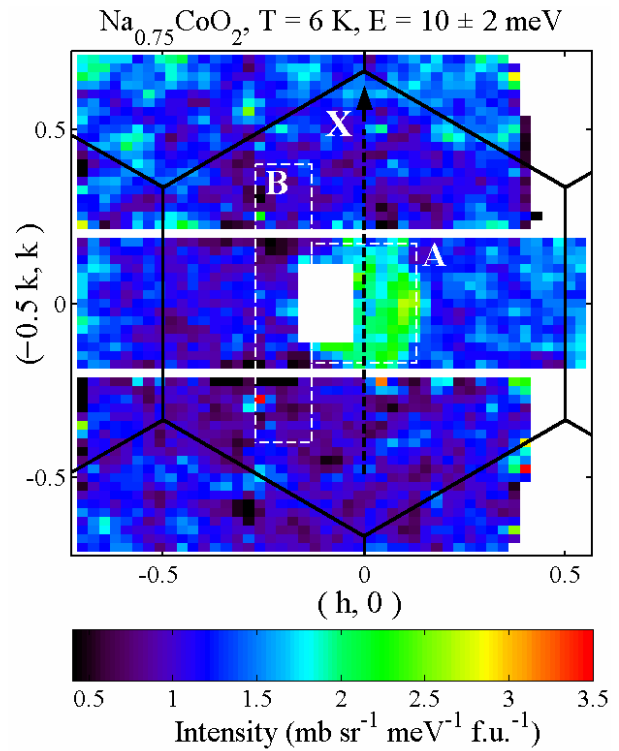}
\caption{(Color online). Neutron inelastic
scattering from Na$_{0.75}$CoO$_2$ recorded in the MAPS
spectrometer area detector with $E_{\rm i}=60$\,meV. The image is
a plot of $(k_{\rm i}/k_{\rm f}){\rm d}^2 \sigma/{\rm d}\Omega{\rm
d}E_{\rm f}$ per formula unit (f.u.), averaged between energies of
8\,meV and 12\,meV, and projected onto the $(h,k)$ plane in
reciprocal space. Blocks of missing data are due to gaps between
detector banks. The hexagonal grid plotted over the data shows the
2D Brillouin zone boundaries. The two rectangular boxes marked A
and B are the areas over which data were averaged for the scans in
Fig.\ \ref{fig:2}. The vertical dotted line marked X indicates the
line along which the cuts shown in Figs.\ \ref{fig:3} and
\ref{fig:4} were taken. \label{fig:1} }
\end{center}
\end{figure}
The physical properties of Na$_x$CoO$_2$ are highly
two-dimensional (2D) \cite{Terasaki-PRB-1997}. It is likely,
therefore, that any magnetic correlations between the CoO$_2$
layers are very weak, and hence that the magnetic scattering is
independent of the component of $\bf Q$ parallel to the crystal
$c$ axis. For an initial survey, therefore, we aligned the crystal
with the $c$ axis parallel to the incident neutron beam, so that
the area detector recorded the energy spectrum over a large region
of 2D reciprocal space ($> 1$ Brillouin zone) spanned by the ${\bf
a}^{\ast}$ and ${\bf b}^{\ast}$ reciprocal lattice basis vectors
of the triangular lattice in the CoO$_2$ plane \cite{RLU}. The
energy range probed in this configuration was from 2\,meV to
50\,meV. By taking a series of constant-energy slices we made a
search of (${\bf Q}_{\rm 2D}$,$E$) space, where ${\bf Q}_{\rm
2D}=h{\bf a}^{\ast}+k{\bf b}^{\ast}\equiv (h,k)$ is the in-plane
component of the scattering vector. The only signal we found that
was clearly in excess of the background in this energy range was
distributed symmetrically around the unscattered beam, i.e.\ ${\bf
Q}_{\rm 2D} = (0,0)$. In particular, there was no observable
signal at wavevectors corresponding to antiferromagnetic
correlations between adjacent Co sites.

To investigate the observed signal further we rotated the crystal
by 30\,degrees, so that more of the ${\bf Q}_{\rm 2D} = (0,0)$
scattering would be recorded in the detectors adjacent to the
unscattered beam. Figure\ \ref{fig:1} displays the neutron
intensity averaged over the energy range 8--12\,meV and projected
onto the $a^{\ast}b^{\ast}$ plane. The unscattered beam passes
through the centre of the blank rectangle where there are no
detectors. The map shows an enhanced signal roughly twice the
background in the detectors to the right of the blank rectangle.
This is the same signal observed in the unrotated orientation, but
now it has shifted to the right due to the rotation of the
crystal. This confirms that the signal is centred on ${\bf Q}_{\rm
2D} = (0,0)$, part of the signal being missing where it overlaps
the blank rectangle.  Away from this feature the intensity
landscape is featureless. We remark here that the peak is much
broader in wavevector than the resolution. The latter is dominated
by the crystal mosaic, which causes a spread in wavevector of
$\sim 0.05 a^{\ast}$ at this energy, whereas the peak has a width
of $\sim 0.2 a^{\ast}$.

\begin{figure}
\begin{center}
\includegraphics
[width=7cm,bbllx=35,bblly=455,bburx=252, bbury=647,angle=0,clip=]
{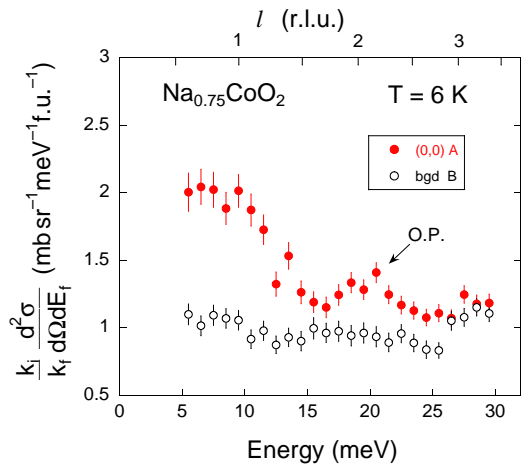} \caption{(Color online). Energy dependence of
the scattering from Na$_{0.75}$CoO$_2$. Filled circles are data
averaged over the rectangular box marked A in Fig.\ \ref{fig:1}
centred on ${\bf Q}_{\rm 2D} = (0,0)$. The background signal has
been estimated from the region marked B in Fig.\ \ref{fig:1}. The
out-of-plane wavevector component $l{\bf c}^{\ast}$ is marked on
the upper axis in reciprocal lattice units (r.l.u.) for the scan
centred on $(0,0)$. The peak marked `O.P.' most likely originates
from an optic phonon.\label{fig:2}}
\end{center}
\end{figure}
Figure\ \ref{fig:2} shows the energy dependence of the scattering
measured at a temperature of 6\,K. To construct this scan we
averaged the data over a rectangular box (marked A in Fig.\
\ref{fig:1}) centred on $(0,0)$ enclosing the peak. Two distinct
features stand out above the background: (1) a peak centred near
20\,meV, and (2) an increase in scattering with decreasing energy
below 15\,meV. The signal appears to level off below 10\,meV, but
this trend cannot be established definitively from the current
data because at low energies a significant proportion of the
signal is lost where there are no detectors, as can be seen in
Fig.\ \ref{fig:1}.

\begin{figure}
\begin{center}
\includegraphics
[width=8cm,bbllx=86,bblly=399,bburx=264, bbury=559,angle=0,clip=]
{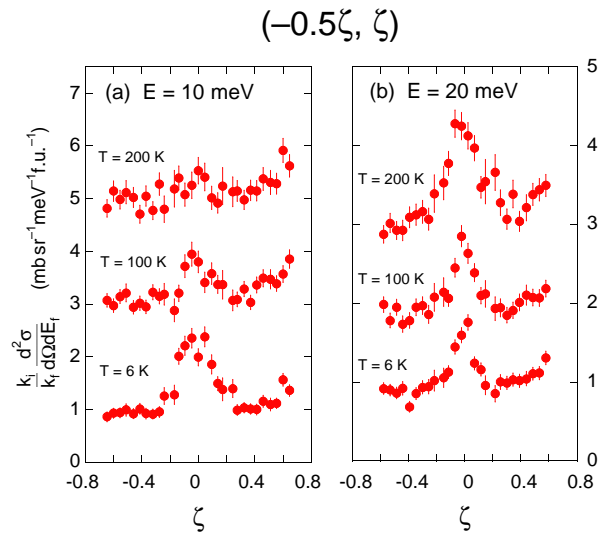} \caption{(Color online). Temperature dependence
of the intensity at ${\bf Q}_{\rm 2D} = (0,0)$ sampled at average
energies of (a) 10\,meV, and (b) 20\,meV. The constant-energy cuts
are taken along the line marked X in Fig.\ \ref{fig:1}. The 100\,K
and 200\,K data have been displaced vertically for clarity. The
displacements relative to the 6\,K data are, respectively, (a) 2
and 3.5 units, and (b) 1 and 2 units. \label{fig:3} }
\end{center}
\end{figure}

Measurements made at higher temperatures revealed that the two
features just described behave very differently with temperature.
The key findings are illustrated in Figs.\ \ref{fig:3}(a) and (b).
The data presented in these plots are cuts through $(0,0)$ along
the line marked X in Fig.\ \ref{fig:1}. The temperature dependence
is shown at average energies of 10\,meV and 20\,meV, respectively.
The peak in the 10\,meV cut is seen to decrease with temperature,
and at 200\,K is almost indistinguishable from the background. By
contrast, the 20\,meV peak increases with temperature.

The increase in intensity with temperature of the 20\,meV peak
suggests that this scattering arises from a bosonic excitation,
such as a phonon. The scattering intensity is then expected to
vary in proportion to the factor $\{1-\exp(-\hbar\omega/k_{\rm
B}T)\}^{-1}$ \cite{Squires}, which increases by $\sim 50$\,\% from
6\,K to 200\,K. This increase is consistent with the data in Fig.\
\ref{fig:3}(b). As there is currently no evidence for magnetic
order at temperatures as high as 200\,K in Na$_{0.75}$CoO$_2$
\cite{Motohashi-PRB-2003} the most likely origin of the 20\,meV
peak is an optic phonon.

On the other hand, the temperature dependence of the low energy
feature suggests a magnetic origin because magnetic correlations
decrease with temperature. The fact that the scattering is
localized about ${\bf Q}_{\rm 2D} = (0,0)$ implies that the
correlations are ferromagnetic within the $ab$ plane.

\begin{figure}
\begin{center}
\includegraphics
[width=5cm,bbllx=43,bblly=410,bburx=212, bbury=651,angle=0,clip=]
{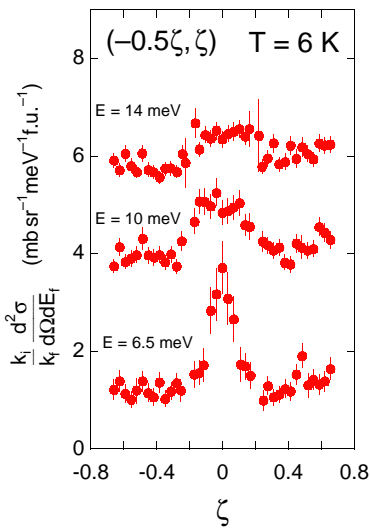} \caption{(Color online). Constant-energy cuts
taken along the line marked X in Fig.\ \ref{fig:1}.
The 10\,meV and 14\,meV data have been displaced vertically by 3
and 5 units, respectively, for clarity. \label{fig:4} }
\end{center}
\end{figure}

Having concluded that the low energy signal at ${\bf Q}_{\rm 2D} =
(0,0)$ corresponds to 2D ferromagnetic correlations we now turn to
its energy spectrum. Fig.\ \ref{fig:4} displays a set of
constant-energy cuts along the line marked X in Fig.\ \ref{fig:1}
through the data collected at $T=6$\,K. The 6.5\,meV cut shows a
single peak centred at $(0,0)$, but with increasing energy the
peak broadens and decreases in amplitude. In addition, the
lineshapes in the 10\,meV and 14\,meV cuts are distinctly
flat-topped, suggestive of two almost-resolved peaks either side
of $\zeta=0$. Although the extent of our data is limited, it is
clear that the magnetic scattering is highly dispersive and that
it extends above $16$\,meV into the energy range where the optic
phonon dominates the signal.

Finally, we consider the energy-integrated cross section, and
compare it with the local-moment sum rule. Integrating the
magnetic signal between 5\,meV and 16\,meV and averaging the
result over one 2D Brillouin zone we obtain $\sim
1$\,mb\,sr$^{-1}$\,f.u.$^{-1}$.
Neglecting the magnetic form factor, the energy-integrated, $\bf
Q$-averaged scattering from a system of local moments with spin
$S$ is $\Sigma = \frac{1}{6}(\gamma r_0)^2g^2S(S+1)$ per spin,
where $\gamma = 1.913$, $r_0 = 2.818 \times 10^{-15}$\,m, and $g$
is the Land\'{e} factor \cite{Squires}. Taking as a reference the
values $S=\frac{1}{2}$ and $g=2$, we find $\Sigma = 145
$\,mb\,sr$^{-1}$\,spin$^{-1}$. In Na$_{0.75}$CoO$_2$ only a
quarter of the Co ions carry a spin, on average, so this would
reduce the expected cross section to $\sim
36$\,mb\,sr$^{-1}$\,f.u.$^{-1}$. This is much larger than the
inelastic scattering cross section we have observed, which
suggests that there is considerable spectral weight outside the
energy range probed in our experiment. Some of this will
undoubtedly be at higher energies, since the measured spectrum
appears to extend above 16\,meV, but we would also expect some of
the weight to be below 2\,meV in the form of elastic or
quasielastic scattering. In particular, although we collected
elastic scattering data over a wide range of reciprocal space we
did not probe along the $(0,0,l)$ line, so any Bragg peaks
associated with in-plane ferromagnetic order could not have been
measured. Full 3D ferromagnetic order is excluded by the
magnetization data \cite{Motohashi-PRB-2003}, but a spin
arrangement with in-plane ferromagnetic order and a spin-density
wave modulation perpendicular to the planes, e.g.\ an
antiferromagnetic stacking along the $c$ axis, would be compatible
with our results.

The measurements reported here show conclusively that there exist
strong ferromagnetic in-plane correlations in Na$_{0.75}$CoO$_2$,
with an energy scale much larger than that set by the observed
magnetic ordering temperature ($T_{\rm m} = 22$\,K
\cite{Motohashi-PRB-2003}). Interestingly, the excitation spectrum
resembles that found in nearly-ferromagnetic metals, such as
Ni$_3$Ga \cite{Bernhoeft-PRL-1989} and Pd \cite{Doubble-PhD-1998},
characterized by strongly damped or overdamped ferromagnons
\cite{Doniach-PPS-1967}. This finding is in remarkable accord with
the picture described by Singh based on density-functional
calculations in the local-density approximation (LDA)
\cite{Singh-PRB-2000,Singh-PRB-2003}. He predicts a weak itinerant
ferromagnetic ground state for all doping levels in the range
$x=0.3-0.7$, and suggests that quantum spin fluctuations would in
practice suppress the tendency for ordering. The spin excitation
spectrum would then resemble that of overdamped ferromagnons,
consistent with what is observed here.

There is good reason to believe that the ferromagnetic
correlations found in Na$_{x}$CoO$_2$ will also be present in the
superconducting compound, and may therefore play a role in the
mechanism of superconductivity. The LDA calculations, which
successfully described the present results, also predict a
ferromagnetic ground state with the strained cell and Na doping of
the hydrated superconductor \cite{Singh-PRB-2003}. This, together
with the evidence reported for unconventional superconductivity
\cite{Unconventional-SC}, raises the possibility of $p$-wave
spin-triplet pairing, a state thought to occur in Sr$_2$RuO$_4$
\cite{APM-RMP-1994}. Unlike the ruthenate, however, the
superconductivity in hydrated Na$_{x}$CoO$_2$ has not been
reported sensitive to impurities, and the most prominent spin
fluctuations in Sr$_2$RuO$_4$ are not ferromagnetic but instead
have an incommensurate wavevector \cite{Sidis_PRL-1999}.
Nevertheless, the proximity to a ferromagnetic phase in each case
suggests the mechanism of superconductivity in the two materials
may share some common features.

We thank Martin Long, Chris Hooley and Andy Mackenzie for
insightful discussions. This work was supported by the Engineering
\& Physical Sciences Research Council of Great Britain.


\begin{references}
\bibitem{RLU}
The reciprocal lattice vectors are defined as ${\bf
a}^{\ast}=\frac{2\pi}{a}\frac{2}{\sqrt{3}}\hat{\bf x}$ and ${\bf
b}^{\ast}=\frac{2\pi}{a}\left( \frac{1}{\sqrt{3}}\hat{\bf x} +
\hat{\bf y} \right)$, where $\hat{\bf x}$ and $\hat{\bf y}$ are
orthonormal vectors and $a=2.83$\,{\AA}
 is the lattice parameter
of the triangular lattice.
\end{references}
\end{document}